\documentclass{elsarticle}

\usepackage{hyperref}

\journal{Journal of \LaTeX\ Templates}

\usepackage{graphicx}

\usepackage{amssymb}
\usepackage{latexsym}
\usepackage{amsmath}

\begin{document}

\begin{frontmatter}

\title{
Two-dimensional transverse-field $XY$ model
with the in-plane anisotropy and Dzyaloshinskii-Moriya interaction:
Anisotropy-driven transition
}

\author{Yoshihiro Nishiyama}
\address{Department of Physics, Faculty of Science,
Okayama University, Okayama 700-8530, Japan}




\begin{abstract}

The two-dimensional
(2D)
quantum spin-$S=1/2$
$XY$ model 
with the 
transverse-field $H$,
in-plane-anisotropy $\gamma$, and
Dzyaloshinskii-Moriya (DM) $D$
interactions
was investigated by means of the exact diagonalization method,
which enables us to treat
the $D$-mediated complex-valued Hermitian matrix elements.
According to the preceding real-space renormalization group analysis at $H=0$,
the $\gamma$-driven phase transition occurs generically for $D\ne 0$ 
in contrast to the 1D $XY$ model where 
both
$\gamma$-  and $D$-induced phases
are realized for
$\gamma>D$ and $ \gamma <D$, respectively.
In this paper, we evaluated the $\beta$ function $\beta(\gamma)$, 
namely, the differential of $\gamma$ with respect to the concerned energy scale,
and 
from its behavior in proximity to $\gamma=0$,
we observed an evidence of the $\gamma$-driven phase transition;
additionally,
$\gamma$'s scaling dimension is estimated from $\beta(\gamma)$'s slope.
It was also determined how the value of the DM interaction influences
the order-disorder phase boundary $H_c(\gamma)$   
around the multi-critical point, $\gamma \to 0$.

\end{abstract}


\end{frontmatter}


\section{\label{section1}Introduction}

\begin{figure}
\includegraphics[width=120mm]{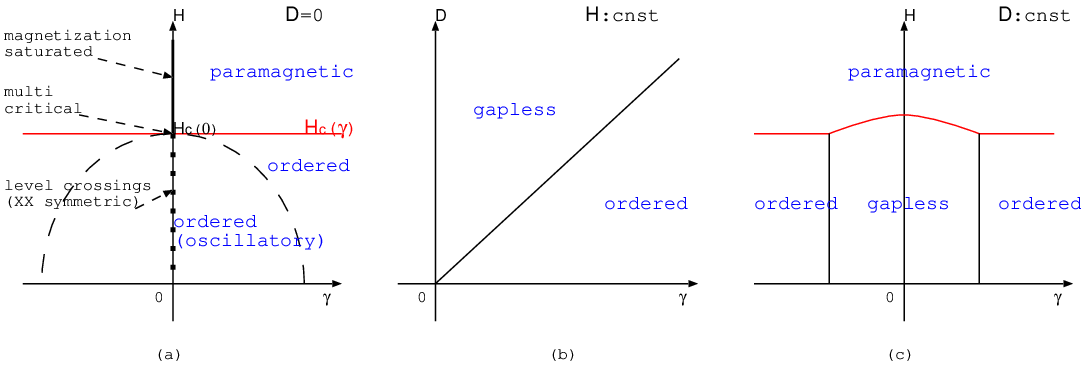}%
\caption{\label{figure1}
(a)
The phase diagram for the one-dimensional transverse-field $XY$ model
with the in-plane anisotropy $\gamma$ and transverse field $H$
is shown \cite{Maziero10,Sun14,Karpat14}.
The phase transition
between the ordered ($H<H_c(\gamma)$)
and paramagnetic ($H>H_c(\gamma)$)
phases
takes place at $H=H_c(\gamma)$.
Along the $\gamma=0$ axis,
owing to the U(1) symmetry,
the intermittent level crossings take place up to $H=H_c(\gamma=0)$
\cite{Mukherjee11},
and the magnetization saturates for the exceedingly large magnetic fields,
	$H>H_c(\gamma=0)$, eventually.
	Within the semicircle (dashed), the correlation function
	gets spatially modulated.
The multi-critical point 
$(\gamma,H)=(0,H_c(0))$
has been investigated extensively \cite{Mukherjee11}.
(b)
the $\gamma$-$D$ phase diagram for the one-dimensional
transverse-field $XY$ model with the DM interaction $D\ne 0$ is shown.
The gapless phase is induced by $D>\gamma$
\cite{Yi19,Ding21,Jafari08}.
(c)
The $\gamma$-$H$ phase diagram under $D\ne 0$ is shown.
In the one-dimensional case, the DM interaction alters the
$\gamma$-$H$ phase diagram significantly
\cite{Yi19,Ding21,Jafari08}.
}
\end{figure}

The one-dimensional $XY$ model 
\begin{equation}
	{\cal H}_{1D}=
- \sum_{i}
((1+\gamma)S^x_{i}S^x_{i+1}+(1-\gamma)S^y_iS^y_{i+1})
-H\sum_{i} S^z_i
	,
\end{equation}
with the
transverse field $H$,
the in-plane anisotropy $\gamma$,
and the spin-$1/2$ operator ${\bf S}_i$ at site $i$
is attracting renewed interest 
\cite{Maziero10,Sun14,Karpat14}
in the context of the quantum information theory
\cite{Luo18,Steane98,Bennett00}.
As shown in Fig. \ref{figure1} (a),
a variety of phases appear in the $\gamma$-$H$ parameter space
\cite{Maziero10,Sun14,Karpat14}.
As the transverse field $H$ changes, there occurs
a phase transition at $H=H_c(\gamma)$,
which separates the ordered ($H<H_c$) and paramagnetic ($H>H_c$) phases.
At the isotropic point, $\gamma=0$, 
because of the U(1) symmetry,
successive level crossings take place
\cite{Mukherjee11} up to $H=H_c(0)$,
whereas
for exceedingly large $H>H_c(0)$, the magnetization is saturated eventually.
Reflecting the level crossings,
the ordered phase with oscillatory correlation function extends 
around the ordinate axis, $H\le H_c(0)$.
Various types of phase boundaries meet 
at the multi-critical point $(\gamma,H)=(0,H_c(0))$,
and this multi-criticality 
has been explored in depth
\cite{Mukherjee11}.
The Dzyaloshinskii-Moriya (DM) interaction $D$ changes the phase diagram significantly
\cite{Yi19,Ding21,Jafari08}.
For instance, the $D$-induced gapless phase appears for sufficiently large $D>\gamma$,
as shown in 
Fig. \ref{figure1} (b).
Correspondingly,
the $\gamma$-$H$ phase diagram 
for $D \ne 0$ exhibits even richer characters \cite{Yi19,Ding21}. 
Actually,
as shown 
in Fig. \ref{figure1} (c),
the phase diagram around the $\gamma=0$ axis is influenced
by $D$.
The DM interaction alters the magnetization saturation point $H_c(0)$ 
as 
\begin{equation}
	\label{saturation_point}
H_c(0)=\sqrt{1+D^2}H_c(0)|_{D=0} 
,
\end{equation}
with the saturation point $H_c(0)|_{D=0}$ for $D=0$
\cite{Yi19,Alcaraz90}. 

\begin{figure}
\includegraphics[width=120mm]{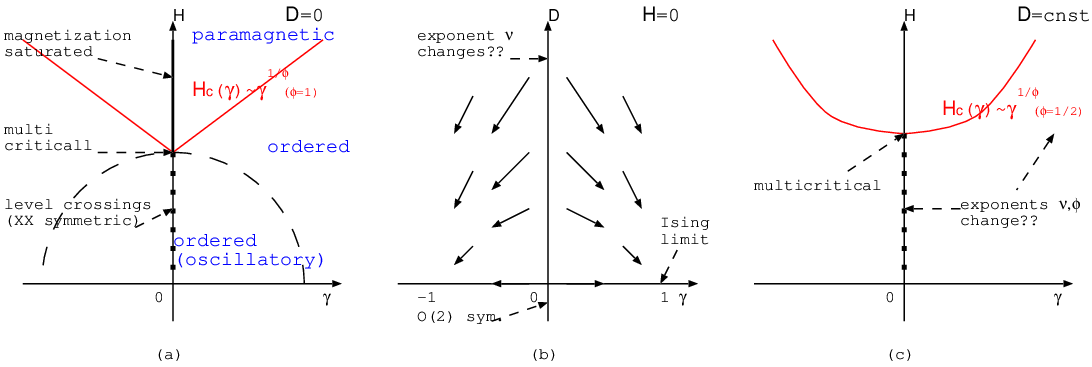}%
\caption{\label{figure2}
(a)
The $\gamma$-$H$ phase diagrams for the
two-dimensional transverse-field $XY$ model
	(\ref{Hamiltonian_original})
with $D=0$ is shown \cite{Jalal16,Wald15,Henkel84,Nishiyama19}.
The
overall features are the same as those of
the one-dimensional counterpart, Fig. \ref{figure1} (a).
A peculiarity of the two-dimensional case
is that the order-disorder phase boundary 
 exhibits
the power-law singularity 
	$H_c(\gamma) -H_c (0)\sim |\gamma|^{1/\phi}$
characterized by
the crossover critical exponent, $\phi=1.00(15)$ 
	[Eq. (\ref{crossover_critical_exponent_D0})]
	\cite{Nishiyama19}.
	Within the semicircle (dashed), the correlation function
	gets spatially modulated.
(b)
The renormalization-group flow \cite{Farajollahpour18} 
in
the $\gamma$-$D$ parameter space at $H=0$ is shown.
The $\gamma$-driven phase transition is suggested,
because the $\gamma=0$ axis is unstable.
(c)
Based on the renormalization-group flow, 
the $\gamma$-$H$ phase diagram with $D\ne 0$ is proposed.
Our concern is to explore
the $\gamma$-driven phase transition \cite{Farajollahpour18}
	as well as $\gamma$'s scaling dimension $1/\nu^{(\gamma)}$.
}
\end{figure}

As for the two-dimensional transverse-field $XY$ model,
the $\gamma$-$H$ phase diagram for $D=0$
has been investigated
both analytically
\cite{Jalal16,Wald15}
and numerically \cite{Wald15,Henkel84,Nishiyama19}.
In Fig. \ref{figure2} (a),
a schematic
$\gamma$-$H$ phase diagram 
obtained with the exact-diagonalization method \cite{Wald15,Henkel84,Nishiyama19}
is shown.
The overall features resemble those of the one-dimensional counterpart, Fig. \ref{figure1} (a);
namely, the $H$-driven phase transition between the ordered and paramagnetic phases
occurs at $H=H_c(\gamma)$.
As for the two-dimensional magnet,
however, the phase boundary $H_c(\gamma)$ 
exhibits a ``monotonous" \cite{Wald15} increase,
as $|\gamma|$ increases.
Such a singularity, namely, the multi-criticality 
at $\gamma=0$,
is characterized
by the crossover critical exponent \cite{Nishiyama19} 
\begin{equation}
\label{crossover_critical_exponent_D0}
	\phi =1.00(15)
,
\end{equation}
which describes the end-point singularity of
the phase boundary
\cite{Riedel69,Pfeuty74} 
as
\begin{equation}
\label{crossover_critical_exponent}
	H_c(\gamma) -H_c(0) \sim |\gamma| ^{1/\phi}
.
\end{equation}
The
phase boundary rises up linearly, $H_c\sim |\gamma|$, in two dimensions,
whereas for the one-dimensional magnet, the phase boundary takes a constant value, $H_c(\gamma)=C$;
see Fig. \ref{figure1} (a) and \ref{figure2} (a).

Meanwhile, 
the DM interaction came under consideration by means of the real-space-renormalization-group 
method at $H=0$ \cite{Farajollahpour18}.
As shown In Fig. \ref{figure2} (b),
the Ising limits $\gamma=\pm 1$ with $D=0$ are the stable renormalization-group
fixed points, and hence,
the $\gamma$-driven phase transition occurs for generic 
values of $D$
in sharp contrast to that of the one dimensional counterpart, Fig. \ref{figure1} (b).
Moreover, it was claimed that $\gamma$'s scaling dimension depends on the DM interaction \cite{Farajollahpour18}.
To the best of author's knowledge, such features have not been studied very extensively by other techniques,
and the $H \ne 0$ case, which include the multi-criticality,
remains totally unclear.

The aim of this paper is to
investigate the $\gamma$-$H$ phase diagram
for the two-dimensional $XY$ model with $D \ne 0$ (see Fig. \ref{figure2} (c)).
We focus our attention on the $\gamma$-driven criticality so as to
examine the real-space-renormalization-group scenario
\cite{Farajollahpour18} for the extended parameter space.
For that purpose,
we employed the
exact diagonalization method,
which enables us to treat the $D$-mediated complex-valued matrix elements.
As a probe to detect the phase transition,
we evaluated the fidelity susceptibility 
\cite{Luo18,Mukherjee11}, which is readily accessible via the
exact diagonalization scheme.
Thereby, we evaluated the fidelity-susceptibility-mediated $\beta$ function, $\beta(\gamma)$,
numerically \cite{Roomany80}.
From its behavior in proximity to the critical point, $\gamma \approx 0$,
we observed an evidence of the $\gamma$-driven phase transition;
we could avoid the complications coming from the level crossings
\cite{Mukherjee11}
along the $\gamma=0$ axis.
It was also determined how
the DM interaction $D$
alters the multi-criticality.


To be specific,
we present the Hamiltonian
for the two-dimensional transverse-field $XY$ model with the DM
interaction
\begin{equation}
\label{Hamiltonian_original}
{\cal H}  =
-J \sum_{\mathbf i} \sum_{{\boldsymbol \delta}={\mathbf e}_{x,y}}(
(1+\gamma)S^x_{\mathbf i}S^x_{{\mathbf i}+{\boldsymbol \delta}}+
(1-\gamma)S^y_{\mathbf i}S^y_{{\mathbf i}+{\boldsymbol \delta}})
-H\sum_{\mathbf i} S^z_{\mathbf i}
	+D\sum_{\mathbf i}  \sum_{{\boldsymbol \delta}={\mathbf e}_{x,y}}
	(
S^x_{\mathbf i}S^y_{{\mathbf i}+{\boldsymbol \delta}}-
S^y_{\mathbf i}S^x_{{\mathbf i}+{\boldsymbol \delta}})
.
\end{equation}
Here, the quantum spin-$S=1/2$
operators $\{ S_{\mathbf i}  \}$ are placed at each square-lattice point ${\mathbf i}$,
and the symbol ${\mathbf e}_{x,y}$ denotes the unit vectors of the lattice.
The parameters, $J$,  $H$, $\gamma$, and $D$, denote
the ferromagnetic nearest-neighbor $XY$ interaction,
the transverse magnetic field, the in-plane anisotropy, and
the DM interaction, respectively;
hereafter, the parameter $J$ is regarded as the unit of energy, {\it i.e.}, $J=1$.
Rather technically, we implemented the screw-boundary condition
\cite{Nakamura23,Nishiyama09,Miyata21} to the finite-size cluster,
and the above expression (\ref{Hamiltonian_original}) has to be remedied accordingly.
The technical details as well as its performance 
are presented
in the next section.
The exact diagonalization method enables us to treat the 
complex-valued 
Hermitian matrix elements due to the DM interaction;
note that 
the $S^y_i$ operator has the pure imaginary matrix elements,
and the exact diagonalization method is free from the sign problem.

It has to be mentioned that the
multi-criticality of the phase boundary $H_c(\gamma)$ has been investigated with the
large-$N$ expansion method \cite{Wald15} for the $d$-dimensional $XY$ model with
$D=0$.
According to this study,
the phase boundary rises up monotonically in large dimensions
$d > 2.065$,
whereas the reentrant behavior, namely, a non-monotonic behavior of $H_c(\gamma)$,
is observed
in the regime, $d< 2.065$.
Therefore, the $d=2$ case locates around the marginal point $d=2.065$,
and it is anticipated that $H_c$'s multi-criticality would be altered by the perturbations
such as the DM interaction.

The rest of this paper is organized as follows.
In the next section, 
the exact numerical 
results are shown.
The above-mentioned screw-boundary condition
\cite{Nakamura23,Nishiyama09,Miyata21}
as well as
its performance check
are presented
in prior to the analysis of the $\gamma$-driven criticality via the $\beta$ function.
In the last section, we present the summary and discussions.

\section{\label{section2}Numerical results}

\begin{figure}
\includegraphics[width=120mm]{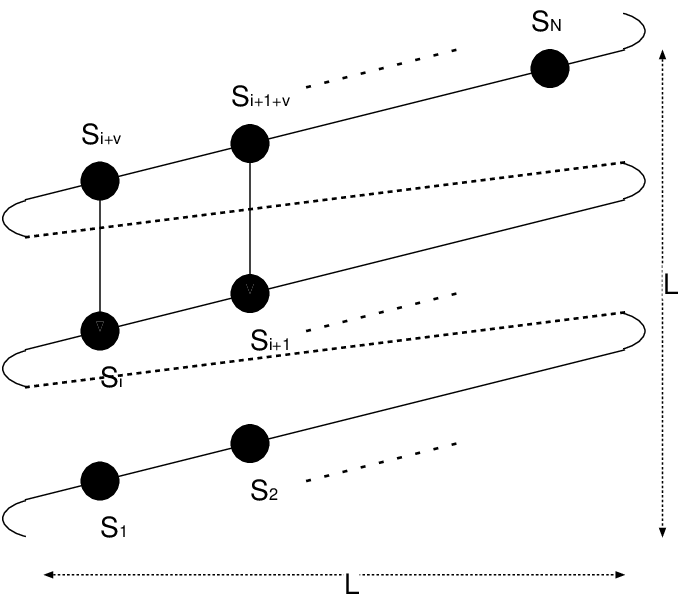}%
\caption{\label{figure3}
The
	screw-boundary condition \cite{Nakamura23,Nishiyama09,Miyata21}
is implemented for the alignment of spins, ${\bf S}_i$
($i=1,2,\dots,N$).
Through the $v(\approx \sqrt{N})$-th neighbor interactions,
the $N$ spins form a 
sheet of network
	effectively.
The Hamiltonian
is given by  Eq. (\ref{Hamiltonian}) explicitly.
	The linear dimension of the sheet  $L$ (\ref{linear_dimensionality})
	is shown. 
}
\end{figure}

In this section, we present the numerical results
for the two-dimensional transverse-field $XY$ model
with the DM interaction.  
We implemented the screw-boundary condition \cite{Nakamura23,Nishiyama09,Miyata21}
for the cluster with $N\le 32$ spins,
as shown in Fig. \ref{figure3}.
To be specific,
the Hamiltonian is given by 
\begin{equation}
\label{Hamiltonian}
{\cal H}=
-J \sum_{i=1}^N \sum_{\delta=1,v}
((1+\gamma)S^x_{i}S^x_{i+\delta}+(1-\gamma)S^y_iS^y_{i+\delta})
-H\sum_{i=1}^{N} S^z_i
+D\sum_{i=1}^N  \sum_{\delta=1,v} (S^x_iS^y_{i+\delta}-S^y_iS^x_{i+\delta})
,
\end{equation}
for an alignment of
spins, 
${\bf S}_i$ with $i=1,2,\dots,N$,
with the screw pitch, $v=[\sqrt{N}+1/2]$.
Here, the bracket $[ \dots ]$ takes
integer part of a number (Gauss notation), {\it e.g.}, $[2.7]=2$,
and the periodic boundary condition, ${\bf S}_{N+1}={\bf S}_1$, is imposed.
As shown in Fig. \ref{figure3},
the $N$ spins form a sheet of network, and
the effective linear dimension of the sheet is given by
\begin{equation}
\label{linear_dimensionality}
L=\sqrt{N}
.
\end{equation}
The linear dimension $L$ plays a significant role
in the scaling analyses such as the $H$-driven criticality in Sec. \ref{section2_1},
where the onset of the Ising universality class is confirmed.

In order to detect the $H$- and $\gamma$-driven phase transitions,
we utilized the fidelity susceptibilities \cite{Luo18,Mukherjee11}, 
$\chi_F^{(H)}$ and 
$\chi_F^{(\gamma)}$, respectively.
To be specific, the former is defined by
\cite{Luo18,Mukherjee11}
\begin{equation}
\label{H-driven_fidelity_susceptibility}
\chi_F^{(H)}=\frac{-1}{N} \partial_{\Delta H}^2 F^{(H)}(H,H+\Delta H)|_{\Delta H =0}
,
\end{equation}
with the fidelity $F^{(H)}(H,H+\Delta H)=|\langle H | H +\Delta H \rangle|$
($\Delta H$: perturbation field), and the ground state $|H\rangle$ of the Hamiltonian
with the magnetic field $H$.
Similarly, the latter type was evaluated 
via
\begin{equation}
\label{gamma-driven_fidelity_susceptibility}
\chi_F^{(\gamma)}=\frac{-1}{N} \partial_{\Delta \gamma}^2 F^{(\gamma)}(\gamma,\gamma+\Delta \gamma)|_{\Delta \gamma =0}
,
\end{equation}
with the fidelity 
$F^{(\gamma)}(\gamma,\gamma+\Delta \gamma)=|\langle \gamma| \gamma+\Delta \gamma)|$, and the ground state 
$|\gamma\rangle$ for the in-plane anisotropy $\gamma$.
The fidelity susceptibility does not rely on any 
{\it ad hoc} assumptions on the order parameters,
{\it i.e.},
either $XX$- ($\gamma=0$) or
Ising-symmetric ($\gamma \ne 0$) one,
and hence,
it is sensitive to generic types of phase transitions.

\subsection{\label{section2_1}Preliminary survey:
The $H$-driven order-disorder transition at the Ising 
$(\gamma,D)=(1,0)$
case
}

As a preliminary survey,
we investigate
the $H$-driven order-disorder phase transition
via the fidelity susceptibility $\chi_F^{(H)}$ (\ref{H-driven_fidelity_susceptibility})
at the Ising point, $\gamma=1$ and $D=0$,
where a number of preceding results are available \cite{Henkel84,Albuquerque10}.
The criticality 
belongs to the {\em classical} three-dimensional (namely, $(2+1)$D) Ising universality class
\cite{Henkel84,Albuquerque10}.

Before showing the 
exact numerical 
results,
we recollect a number of scaling relations for $\chi_F^{(\lambda)}$
($\lambda =H,\gamma$).
In general \cite{Albuquerque10},
the fidelity susceptibility obeys
the scaling formula
\begin{equation}
\label{scaling_formula}
\chi_F^{(\lambda)}= L^x f( (\lambda-\lambda_c)L^{1/\nu}   )
,
\end{equation}
with the critical point $\lambda_c$,
and
the correlation-length critical exponent $\nu$.
Here,
$\chi_F^{(\lambda)}$'s scaling dimension 
$x$
is given by
\cite{Albuquerque10}
\begin{equation}
\label{scaling_relation}
x= \alpha/\nu +z
,
\end{equation}
with
the specific-heat critical exponent $\alpha$,
and
the dynamical critical exponent
$z$.
Therefore,
as for the classical three-dimensional Ising universality class,
the scaling dimension takes the value
\begin{equation}
\label{3DI_scaling_dimension}
x_{3DI}=1.1739
,
\end{equation}
through resorting to the
critical exponents,
$\nu_{3DI}=0.63012$, $\alpha_{3DI}=0.1096$ \cite{Campostrini02}, and 
$z_{3DI}=1$ \cite{Henkel84,Albuquerque10}.
The critical exponents appearing in the scaling formula
(\ref{scaling_formula})
are all fixed,
and we are able to
analyze the order-disorder phase transition
via $\chi_F^{(H)}$ (\ref{H-driven_fidelity_susceptibility}).

\begin{figure}
\includegraphics[width=120mm]{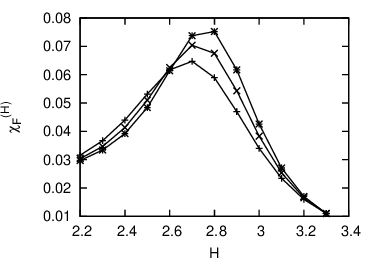}%
\caption{\label{figure4}
The fidelity susceptibility
$\chi_F^{(H)}$ (\ref{H-driven_fidelity_susceptibility}) is plotted for various $H$ and system sizes
($+$) $N=24$,
($\times$) $28$, and
($*$) $32$
with the fixed $\gamma=1$ and $D=0$ (Ising limit).
The peak around $H =H_c \approx 2.8$ indicates an onset of the phase transition,
which separates the ordered ($H<H_c$) and paramagnetic
($H>H_c$) phases;
see Fig. \ref{figure2} (a).
	}
\end{figure}

In Fig. \ref{figure4},
we present
the fidelity susceptibility $\chi_F^{(H)}$ 
(\ref{H-driven_fidelity_susceptibility})
for various $H$ and $N=24$-$32$ with the fixed 
$\gamma=1$ and $D=0$ (Ising limit).
The fidelity susceptibility $\chi_F^{(H)}$ exhibits a notable peak around 
$H \approx 2.8$,
which indicates an onset of the phase transition 
between
the ordered ($H<H_c$) and paramagnetic ($H>H_c$)
phases;
see the phase diagram in Fig. \ref{figure2} (a).

\begin{figure}
\includegraphics[width=120mm]{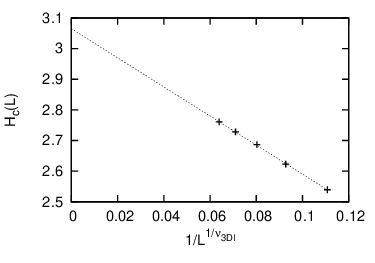}%
\caption{\label{figure5}
The approximate critical point
	$H_c(L)$ (\ref{approximate_critical_point})
	is plotted for $1/L^{1/\nu_{3DI}}$ with $\nu_{3DI}=0.63$ 
\cite{Henkel84,Albuquerque10}.
The parameters, $\gamma=1$ and $D=0$, are the same as those of Fig. \ref{figure4}.
The least-squares fit to these data yields an estimate $H_c=3.065(4)$ in the thermodynamic limit
$L\to \infty$.
Possible systematic error is considered in the text.
}
\end{figure}

In order to estimate the critical point $H_c$ precisely,
in Fig. \ref{figure5},
the approximate critical point $H_c(L)$ is plotted for $1/L^{1/\nu_{3DI}}$ with the correlation-length critical exponent
$\nu_{3DI}=0.63$ \cite{Henkel84,Albuquerque10}.
The data should align,  
because the argument of the function $f$ in Eq. (\ref{scaling_formula}) is a scale-independent constant
$(H-H_c)L^{1/\nu_{3DI}}=C$, which indicates $H_c(L)= C (\frac{1}{L^{1/\nu_{3DI}}})+H_c$.
Here, the parameters, $\gamma=1$ and $D=0$, are the same as those of Fig. \ref{figure4}.
The approximate critical point $H_c(L)$ denotes
$\chi_F^{(H)}$'s peak position
\begin{equation}
\label{approximate_critical_point}
\partial_H \chi_F^{(H)}|_{H=H_c(L)}=0
,
\end{equation}
for each system size $L(=\sqrt{N})$ (\ref{linear_dimensionality}).
The least-squares fit to these data yields an estimate 
$H_c=3.065(4)$
in the thermodynamic limit $L\to\infty$.
The data in Fig. \ref{figure5} appear to be convexly curved because of the corrections to the finite-size scaling
\cite{Challa86b}
as well as
the screw-boundary condition  \cite{Novotny90}.
Actually, in the screw-boundary condition \cite{Novotny90},
a wide range of $N=16,20,\dots,32$ has to be considered,
because the wavy deviation between the quadratic $N=4^2$, $5^2$ and an intermediate $N=20(\approx 4.5^2)$
appears, and such an undulation has to be smeared out by including a sector of $N=4^2$-$5^2$ at least.
On the one hand, as shown in Fig. 8 of Ref. \cite{Challa86b}, the approximate critical point $H_c(L)$ is curved
convexly, and the extrapolated critical point $H_c$ for the largest $N=24,28,32$ drifts to a slightly smaller value. 
In order to appreciate the amount of errors,
we made the similar  analysis for the largest three system sizes, $N=24,28,32$,
which do not contain $N=4^2$.
As a result, we arrive at 
an estimate $H_c=3.048(1)$; the deviation from the above, $\approx 0.02$ seems to dominate
the least-squares-fit error, $\approx 0.004$; corrections to scaling
appear to be non-negligible.
Hence, considering the former as the main source of uncertainty,
we estimate the critical point
as
\begin{equation}
\label{critical_point}
H_c=3.065(20)
.
\end{equation}
Our result (\ref{critical_point}) appears to agree with
the quantum Monte Carlo $H_c=3.0442(4)$
\cite{Albuquerque10}
and exact diagonalization 
$3.05(1)$
\cite{Henkel84} results,
confirming the validity of the 
numerical scheme based on
the screw-boundary condition, Eq (\ref{Hamiltonian}).


\begin{figure}
\includegraphics[width=120mm]{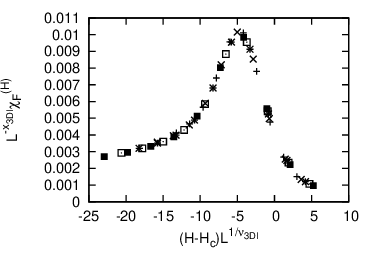}%
\caption{\label{figure6}
The scaling plot,
	$(H-H_c)L^{1/\nu_{3DI}}$-$L^{-x_{3DI}}\chi_F^{(H)}$, is shown for 
	($+$) $N(=L^2)=16$,
	($\times$) $20$, 
	($*$) $24$,
	($\Box$) $28$, and
	($\blacksquare$) $32$
	with $H_c=3.065$ (\ref{critical_point}),
	$\nu_{3DI}=0.63$ \cite{Henkel84,Albuquerque10},
	and $x_{3DI}=1.1739$ (\ref{3DI_scaling_dimension}),
	based on the scaling formula (\ref{scaling_formula});
	the parameters $\gamma=1$ and $D=0$ (Ising limit) are the same as those of Fig. \ref{figure4}.
The scaled data collapse into a scaling curve satisfactorily.
}
\end{figure}

As a cross-check,
in Fig. \ref{figure6},
we present the scaling plot,
$(H-H_c)L^{1/\nu_{3DI}}$-$L^{-x_{3DI}}\chi_F^{(H)}$, 
for various system sizes, $N=16$-$32$,
with $H_c=3.065$ (\ref{critical_point}),
$\nu_{3DI}=0.63$ \cite{Henkel84,Albuquerque10},
and 
$x_{3DI}=1.1739$
(\ref{3DI_scaling_dimension}),
based on the finite-size scaling formula (\ref{scaling_formula}).
The scaled data fall into the scaling curve satisfactorily,
confirming the validity of our 
exact numerical
scheme.
We stress that
the scaling parameters, $H_c$, $\nu_{3DI}$, and
$x_{3DI}$, were all fixed in prior to the scaling analysis,
and there is no {\it ad hoc} adjustable parameter 
in the analysis of Fig. \ref{figure6}.
The scaled peak position in Fig. \ref{figure6} locates around the
off-critical regime, $(H-H_c)L^{1/\nu_{3DI}}\approx -5$.
In the thermodynamic limit
$L\to\infty$, 
this relation indicates that the peak position converges to the critical point
as
$H (= H_c -5/L^{1/\nu_{3DI}}) \to H_c$.

\subsection{\label{section2_2}
The $\gamma$-driven criticality for $D\ne 0$ and $H=0$}

In this section, we analyze the $\gamma$-driven phase transition for $D \ne0$ and $H=0$
(see Fig. \ref{figure2} (b)) with the aid of  the fidelity susceptibility, $\chi_F^{(\gamma)}$
(\ref{gamma-driven_fidelity_susceptibility}).
As mentioned in Introduction, we evaluated the $\beta$ function \cite{Roomany80}
\begin{equation}
\label{beta_function}
	\beta(\gamma,N)=
	\frac{
	x^{(\gamma)}-
	\log(\chi_F^{(\gamma)}(N)/\chi_F^{(\gamma)}(N-4))/\log(\sqrt{N/(N-4)})
}{
	\sqrt{
\partial_\gamma \chi_F^{(\gamma)}(N) 
\partial_\gamma \chi_F^{(\gamma)}(N-4)/\chi_F^{(\gamma)}(N) /\chi_F^{(\gamma)}(N-4)
		}
}
\end{equation}
with the fidelity susceptibility $\chi_F^{(\gamma)}(N)$ (\ref{gamma-driven_fidelity_susceptibility})
for the system size $N$,
and $\chi_F^{(\gamma)}$'s scaling dimension $x^{(\gamma)}$ (\ref{scaling_relation}).
Formally, the $\beta$ function is defined by the differential
of the coupling constant $\gamma$ with respect to the concerned energy scale
\begin{equation}
	\label{formal_beta_function}
	\beta(\gamma)=\frac{{\rm d}\gamma}{{\rm d}\ln \Lambda}
	,
\end{equation}
with the cut-off $\Lambda$.
This formal expression is well approximated by the above 
formula (\ref{beta_function}) \cite{Roomany80}.
Because the argument of the function $f$ in Eq. (\ref{scaling_formula})
is a constant $(\gamma-\gamma_c)L^{1/\nu^{(\gamma)}}=C$,
and the inverse of the system size $L$ sets the cut-off $\Lambda \sim 1/L$,
the formal expression (\ref{formal_beta_function})
reduces to
\begin{equation}
	\label{beta_function_asymptotic_form}
	\beta (\gamma) = \frac{1}{\nu^{(\gamma)}}  (\gamma-\gamma_c)
	,
\end{equation}
with the correlation-length critical exponent $\nu^{(\gamma)}$
and the transition point, $\gamma = \gamma_c$.
Therefore, $\gamma$'s scaling dimension, $1/\nu^{(\gamma)}$, is estimated from the
slope of $\beta(\gamma)$ in the vicinity of the critical point, $\gamma \to \gamma_c$.
According to the real-space normalization group \cite{Farajollahpour18},
the $\gamma$-driven transition should take place at
\begin{equation}
	\label{gamma-driven_critical_point}
\gamma_c=0
.
\end{equation}
We stress that the asymptote (\ref{beta_function_asymptotic_form})
is realized 
in close vicinity of the critical point,
$\gamma \to 0$. 
Actually, the power-law singularity of the critical phenomenon is well defined
in proximity to the critical point,
and hence, the slope of the $\beta$ function for $\gamma \to \gamma_c$,
namely, the first derivative of $\beta(\gamma)$ around the critical point,
captures the concerned critical exponent $\nu^{(\gamma)}$
correctly.
The 
numerical data suffer from the finite-size artifact \cite{Mukherjee11}
due to the level crossings along the $\gamma=0$ axis; see Fig. \ref{figure2} (a).
Therefore, from the proximate behavior of the $\beta$ function beside $\gamma \approx 0$,
we observe the critical behavior (particularly, $\beta(\gamma)$'s slope, $1/\nu^{(\gamma)}$),
avoiding the complications arising from the level crossings at $\gamma=0$.
In order to evaluate
the $\beta(\gamma)$ via Eq. (\ref{beta_function}),
we need to fix the 
the scaling dimension $x^{(\gamma)}$.
Putting $\alpha/\nu(=d)=2$ ($d$: dimensionality)
\cite{Challa86b} and $z=1$ (same as that of Sec. \ref{section2_1})
into the scaling relation (\ref{scaling_relation}),
we obtain
\begin{equation}
x^{(\gamma)}=3
.
\end{equation}
We put this value into the 
the $\beta$-function formula (\ref{beta_function}),
and evaluated it by the exact numerical method.

\begin{figure}
\includegraphics[width=120mm]{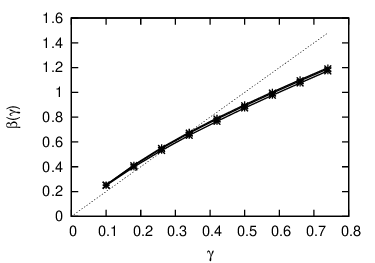}%
\caption{\label{figure7}
The $\beta$ function, $\beta(\gamma)$ (\ref{beta_function}), is plotted for 
various $\gamma$, and
	($+$) $D=0$
	($\times$) $0.3$,
	and
	($*$) $0.5$ with the fixed $H=0$ and $N=32$.
	The $\beta$ function seems to obey the
	formula, $\beta(\gamma)=(\gamma-\gamma_c)/\nu^{(\gamma)}$
	(\ref{beta_function_asymptotic_form}),
	with the slope $1/\nu^{(\gamma)}=2$ (\ref{gamma-driven_critical_exponent})
	and 
	the transition point $\gamma_c=0$ (\ref{gamma-driven_critical_point})
	for sufficiently small $\gamma$,
	as shown by the dotted line, $\beta(\gamma)=2\gamma$.
	The critical exponent $\nu^{(\gamma)}=1/2$ (\ref{gamma-driven_critical_exponent}) seems to be retained
	even for $D \ne 0$.
	The 
	exact numerical calculation
	in close vicinity of $\gamma=0$ fails
	because of the level crossings \cite{Mukherjee11} at $\gamma=0$.
	Our result is summarized in Table \ref{table}.
}
\end{figure}

In Fig. \ref{figure7},
we present the $\beta$ function,
$\beta(\gamma)$ (\ref{beta_function}),
for various $\gamma$
and 
($+$) $D=0$,
($\times$) $D=0.3$, and
($*$) $D=0.5$
with the fixed $H=0$ and $N=32$.
We also show the line $\beta(\gamma)=2\gamma$ as a dotted line.
The numerical data approach to this line asymptotically for sufficiently small $\gamma$.
According to Eq. (\ref{beta_function_asymptotic_form}),
the slope of this line and the $\gamma$-intercept yield 
$1/\nu^{(\gamma)}$ and $\gamma_c$, respectively.
Hence, we estimate the correlation-length critical exponent
\begin{equation}
\label{gamma-driven_critical_exponent}
\nu^{(\gamma)}=1/2
,
\end{equation}
and the transition point $\gamma_c=0$ (\ref{gamma-driven_critical_point}).
The critical exponent
$\nu^{(\gamma)}=1/2$
(\ref{gamma-driven_critical_exponent})
is identical to that of Fig. 6 of Ref. \cite{Binder84},
where the scaling dimension of the symmetry-breaking magnetic field $h$, 
$1/\nu^{(h)}=2$, 
and the critical point, $h_c=0$, are  estimated 
for the Ising model.
Therefore, 
regarding $\gamma$ as the symmetry-breading field $h$ in
Ref. \cite{Binder84},
the underlying physics is the same as ours,
and
the renormalization-group result
$\gamma_c=0$ (\ref{gamma-driven_critical_point}) \cite{Farajollahpour18}
is supported by this preceding result \cite{Binder84}.
Here, we stress that this idea is retained even in the presence 
of the DM interaction, because the numerical data in Fig. \ref{figure7} are almost
independent on $D$;
this is not so trivial, as shown in the analysis of the multi-criticality  in Sec. \ref{section2_4}.
Moreover, 
the deviation of the numerically evaluated $\beta$ function
from the asymptote 
was observed in
the left panel of Fig. 11 of Ref. \cite{Binder84},
where
the logarithmic plot for the magnetic susceptibility is shown, and 
the anticipated slope is realized only within a narrow window beside the critical point $h_c=0$.
The situation of Ref. \cite{Binder84} is essentially
the same as the $\beta$ function,
where the logarithmic discrete derivative of $\chi_F^{(\gamma)}$ is taken as shown in Eq. (\ref{beta_function}).
Hence, substantial improvement of the convergence is only attained by enlarging $N$
at a geometrical rate, and such a treatment cannot be managed by the numerical methods.
The small-$\gamma<0.1$ results for $\beta(\gamma)$ are missing because of the following reason.
At the rotational symmetric point $\gamma=0$, the total magnetization $M^z=\sum_{i=1}^N S^z_i$
commutes with the Hamiltonian, and it takes the quantized values, $M^z=-N/2,-N/2+1,\dots,N/2$.
Therefore,
as the conjugate magnetic field $H$ changes,
the magnetization shows intermittent jumps because of the successive level crossings;
see Fig. \ref{figure2} (a).
Owing to the level crossings, the numerical results in close vicinity of $\gamma < -0.1$ get
scattered, and those results are missing in Fig. \ref{figure7}.

We address a number of remarks.
First,
as shown in Table \ref{table},
the case of $H=0$ and $D=0$ has been investigated
rather extensively by means of 
the real-space-renormalization-group method \cite{Usman16,Farajollahpour18}.
In Ref. \cite{Usman16},
an information-theoretical quantifier, the so-called concurrence $C$, was 
evaluated, and from $\partial_\gamma C$'s peak position $\gamma_{max}$, and the peak height
$\partial_\gamma C|_{\gamma=\gamma_{max}}$, the critical exponent was estimated as $\nu^{(\gamma)}=1.14^{-1}$
and $1.35^{-1}$, respectively.
Furthermore, the result
$\nu^{(\gamma)}=0.4869$ was obtained from the power-law singularity of the energy gap
\cite{Farajollahpour18}.
The latter is in perfect agreement
with ours $\nu^{(\gamma)}=1/2$ (\ref{gamma-driven_critical_exponent}).
%
%
Second,
the case of $D \ne 0$ and $H=0$ has been investigated
in Ref. \cite{Farajollahpour18}.    
According to this elaborated study \cite{Farajollahpour18}, the critical exponent $\nu^{(\gamma)} (D)$
should decrease, as $|D|$ increases.
Our exact numerical 
result indicates that the critical exponent $\nu^{(\gamma)}=1/2$ (\ref{gamma-driven_critical_exponent})
is retained even for the non-zero $D$. 
Last,
we stress that 
the slope of the $\beta$ function in the vicinity of the critical point
makes sense.
We recall the correlation-length power-law divergence $\xi \sim |\gamma-\gamma_c|^{-\nu^{(\gamma)}}$
\cite{Vojta03}.
From the power-law divergence of $\xi$, the 
expression
	$\beta (\gamma) =   (\gamma-\gamma_c)/\nu^{(\gamma)}$
	(\ref{beta_function_asymptotic_form})
	is derived through resorting to
	the formal definition
	$\beta(\gamma)=\frac{{\rm d}\gamma}{{\rm d}\ln \Lambda}$
	(\ref{formal_beta_function}), and $\Lambda(\sim 1/L)\sim 1/\xi$.


\begin{table}[t]
\centering
\begin{tabular}{l l l l}
	method & $\nu^{(\gamma)}|_{D=0,H=0}$ (probe) & $\nu^{(\gamma)}(D)|_{H=0}$ & $\dot{\nu}^{(\gamma)}(D)$ \\ 
\noalign{\smallskip}\hline\noalign{\smallskip}
	RSRG \cite{Usman16} & $1.35^{-1}$ ($\frac{\partial C}{\partial \gamma}|_{\gamma_{max}}$), $1.14^{-1}$ ($\gamma_{max}$) &  & \\
	RSRG \cite{Farajollahpour18} & $0.4869$ (energy gap) & decreases &  \\
ED	(this work) & $\frac{1}{2}$ ($\beta(\gamma)$'s slope) & $\frac{1}{2}$ & $1$ \\
\end{tabular}
\caption{
	So far, the $\gamma$-driven correlation length critical exponent
	$\nu^{(\gamma)}$ has been estimated by means  of the real-space renormalization group 
	(RSRG) method
	\cite{Usman16,Farajollahpour18} for a number of limiting cases.
	In Ref. \cite{Usman16}, the information theoretical quantifier,
	the so-called concurrence $C$, was analyzed, whereas in Ref. \cite{Farajollahpour18},
	the power-law behavior of the energy gap was studied.
	Our exact diagonalization (ED) results are also shown.
	}\label{table}
\end{table}

\subsection{\label{section2_3}
The $\gamma$-driven criticality for $D\ne0$ and $H\ne 0$}

In this section, we investigate the case of $D \ne 0$
and $H \ne 0$, which is not covered by the preceding studies \cite{Usman16,Farajollahpour18}.

\begin{figure}
\includegraphics[width=120mm]{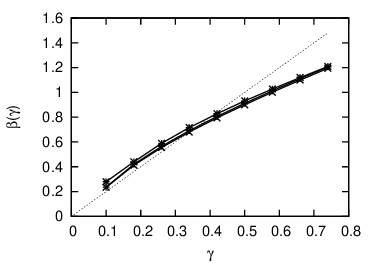}%
\caption{\label{figure8}
The $\beta$ function, $\beta(\gamma)$ (\ref{beta_function}), is plotted for 
various $\gamma$, and
	($+$) $(D,H)=(0.3,0.5)$
	($\times$) $(0.5,0.5)$,
	and
	($*$) $(0.7,1)$ with $N=32$.
	The asymptote
	$\beta(\gamma)=\gamma/\nu^{(\gamma)}$
	(\ref{beta_function_asymptotic_form})
	with the slope $1/\nu^{(\gamma)}=2$ (\ref{gamma-driven_critical_exponent})
	is shown as
	the dotted line, $\beta(\gamma)=2\gamma$.
	The critical exponent $1/\nu^{(\gamma)}=2$ seems to be retained
	for $ D \ne 0$ and  $H \ne 0$.
	The data of ($*$) $(D,H)=(0.7,1)$ show a 
	slight deviation from
	$\beta(\gamma)=2\gamma$,
	and such a regime of large $D$ and $H$ is considered 
	in the subsequent analysis in regard to the multi-criticality.
}
\end{figure}

In Fig. \ref{figure8},
we present the $\beta$ function, $\beta(\gamma)$ (\ref{beta_function}), for various $\gamma$,
and 
($+$) $(D,H)=(0.3,0.5)$,
($\times$) $(0.5,0.5)$, and
($*$) $(0.7,1)$ with $N=32$.
The $\beta$ function seems to obey the asymptotic form,
$\beta(\gamma)=\gamma/\nu^{(\gamma)}$ 
(\ref{beta_function_asymptotic_form}),
with the slope $1/\nu^{(\gamma)}=2$
(\ref{gamma-driven_critical_exponent})
for sufficiently small $\gamma < 0.3$,
as in the case of  $H = 0$ (Sec. \ref{section2_2}).
As mentioned in Sec. \ref{section2_2}, the
critical exponent $1/\nu^{(\gamma)}=2$
(\ref{gamma-driven_critical_exponent})
is identical to that of the symmetry-breaking-field-driven 
criticality $1/\nu^{(h)}=2$ as in Ref. \cite{Binder84}, regarding
the anisotropy $\gamma$ as the symmetry breaking field.
Hence, the underlying physics is the same as ours,
and
we stress that the idea is validated even
in the presence of $H$ as well as $D$.
Again, the deviation of the numerically evaluated $\beta$ function
from the asymptote should be attributed to the subtlety of the discrete logarithmic derivative 
of $\chi_F^{(\gamma)}$ in Eq. (\ref{beta_function}), as argued in Sec. \ref{section2_2}.
Taking a closer look,
we notice that for 
rather large 
$(D,H)=(0.7,1)$,
the results 
show a slow convergence to the asymptote $\beta(\gamma)=2\gamma$. 
Such a feature for
large-$(d,H)$ regime should be regarded as a precursor of the multi-criticality,
which is studied in the next section.

We address a number of remarks.
First,
according to 
the renormalization-group analysis for 
$H=0$ \cite{Farajollahpour18},
the
critical exponent
$\nu^{(\gamma)}(D)$ decreases, as $D$ increases.
In contrast, the present analysis 
suggests that the critical exponent
$\nu^{(\gamma)}=1/2$ (\ref{gamma-driven_critical_exponent})
is robust against the variation of $H$ as well as $D$.
Last, our results in Sec. \ref{section2_2} and \ref{section2_3} indicate that the $\gamma$-driven phase transition 
occurs  at $\gamma_c=0$ (\ref{gamma-driven_critical_point}).
Such a feature provides a marked contrast to that of the one-dimensional counterpart (Fig. \ref{figure1} (c)),
where a transient gapless phase is induced by $\gamma$ for $D \ne 0$.
A peculiarity of the one-dimensional $XX$ ($\gamma=0$)
model is that the magnetic order develops only marginally (gapless), and 
in the presence of the DM interaction,
the in-plane anisotropy $\gamma$ 
cannot support the magnetic order.
On the contrary, in two dimensions, the long-range order develops
for the $XX$-symmetric ($\gamma=0$) case, and
the non-zero $\gamma$ term immediately leads to 
the magnetic order along the easy-axis direction even in the presence of $D$.

\subsection{\label{section2_4}
The $\gamma$-driven criticality for large $H(\approx H_c(0))$:
Multi-criticality for $D \ne 0$}

In this section, we show an evidence that 
for large $H (\approx  H_c(0))$ (\ref{saturation_point}),
the $\gamma$-driven critical exponent becomes
\begin{equation}
\label{multi-critical_critical_exponent}
	\dot{\nu}^{(\gamma)}=1
,
\end{equation}
instead of $1/\nu^{(\gamma)}=2$ (\ref{gamma-driven_critical_exponent}) eventually;
see Table \ref{table}.
This result indicates that
owing to $D \ne 0$,
the crossover critical exponent $\phi$ changes,
and accordingly,
the phase boundary $H_c(\gamma) \sim |\gamma|^{1/\phi}$ (\ref{crossover_critical_exponent})
becomes curved quadrically, as shown in Fig. \ref{figure2} (c).

Before commencing the analysis of the multi-critical exponent $\dot{\nu}^{(\gamma)}$
via $\beta(\gamma)$,
we recollect related multi-critical scaling relations so as to
elucidate the implications of $\dot{\nu}^{(\gamma)}=1$ (\ref{multi-critical_critical_exponent}).
The crossover critical exponent $\phi$ (\ref{crossover_critical_exponent}) is given by  \cite{Riedel69,Pfeuty74}
\begin{equation}
\phi=\dot{\nu}^{(H)} / \dot{\nu}^{(\gamma)}
,
\end{equation}
with the
$H$- and $\gamma$-driven 
correlation-length critical exponents, $\dot{\nu}^{(H)}$ and $\dot{\nu}^{(\gamma)}$, respectively,  at the multi-critical point 
$H=H_c(0)$ (\ref{saturation_point}).
As for 
$\dot{\nu}^{(H)}$,
we set
\begin{equation}
	\dot{\nu}^{(H)} =1/2
	,
\end{equation}
which describes 
the $H$-induced phase transition to the fully-polarized state
\cite{Zapf14}.
Hence, combining this with the aforementioned one, $\dot{\nu}^{(\gamma)}=1$ (\ref{multi-critical_critical_exponent}),
we arrive at
\begin{equation}
	\label{D_crossover_critical_exponent}
\phi=1/2
,
\end{equation}
for $D \ne 0$.
This result indicates that 
the phase boundary should
curve
quadratically, $H_c\sim |\gamma|^2$
(\ref{crossover_critical_exponent}), as mentioned in the first paragraph of this section.

\begin{figure}
\includegraphics[width=120mm]{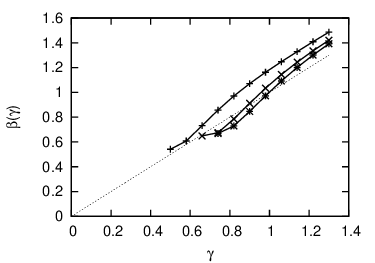}%
\caption{\label{figure9}
The $\beta$ function, $\beta(\gamma)$ (\ref{beta_function}), is plotted for 
various $\gamma$, and
	($+$) $(D,H)=(0.6,2.3)$
	($\times$) $(0.7,2.45)$,
	and
	($*$) $(0.8,2.5)$ with $N=32$.
	The asymptote
	$\beta(\gamma)=\gamma/\dot{\nu}^{(\gamma)}$
	(\ref{beta_function_asymptotic_form})
	with the slope $1/\dot{\nu}^{(\gamma)}=1$ (\ref{multi-critical_critical_exponent})
	is shown by
	the dotted line, $\beta(\gamma)=\gamma$.
}
\end{figure}

In Fig, \ref{figure9},
we present the $\beta$ function, $\beta(\gamma)$ (\ref{beta_function}), for various $\gamma$,
and 
($+$) $(D,H)=(0.6,2.3)$,
($\times$) $(0.7,2.45)$, and
($*$) $(0.8,2.5)$ with $N=32$.
The $\beta$ function appears to obey the 
asymptotic form,
$\beta(\gamma)=\gamma/\dot{\nu}^{(\gamma)}$ 
(\ref{beta_function_asymptotic_form}),
with the slope $1/\dot{\nu}^{(\gamma)}=1$
(\ref{multi-critical_critical_exponent}), as indicated by the dotted line.
The deviation of the numerically evaluated $\beta$ function from the asymptote
was observed in
the right panel of Fig. 11 of Ref. \cite{Binder84},
where
the logarithmic plot of the susceptibility for the Ising model at the critical end-point 
is shown, and 
the anticipated slope is realized only within an extremely narrow window.
As shown in Ref. \cite{Binder84},
Such an end-point singularity (multi-criticality at $H=H_c(0)$ is affected by
the criticality along the branch ($\gamma=0$ and $H< H_c(0)$) in a transient manner
for finite system sizes, because
the former slope
$1/ \dot{\nu}^{(\gamma)}=1$
is smaller than the latter 
$1/\nu^{(\gamma)}=2$.
Moreover,
the $\dot{\nu}^{(\gamma)}=1$
(\ref{multi-critical_critical_exponent}) 
is mathematically supported, because this exponent appears
in the exact solution for the $d=1$ spin chain 
(see Fig. \ref{figure1} (c));
note that the exponent $\dot{\nu}^{(\gamma)}=1$
means $H_c(\gamma)\sim |\gamma|^2$,
which is derived by the rigorous argument.
(The fractional (non-integral) value of $\dot{\nu}^{(\gamma)}$
yields the power-law singularity as to $H_c(\gamma)$, and such a singularity
is not validated by the rigorous argument \cite{Wald15}.)
Taking a closer look, rather small-$D=0.6$ data exhibit a slow convergence to the
asymptote, suggesting that the $D$ term is significant to realize $\dot{\nu}^{(\gamma)}=1$.



We stress that the multi-critical exponent 
$\dot{\nu}^{(\gamma)}=1$
(\ref{multi-critical_critical_exponent})
right at $H=H_c(0)$
differs significantly from
$\nu^{(\gamma)}=1/2$
(\ref{gamma-driven_critical_exponent}) for the $|H|< H_c(0)$ branch:
Actually,
as shown
in Fig. \ref{figure9}, for sufficiently large ($\times$) $D=0.7$ and ($*$) $0.8$,
the 
exact numerical results are almost overlapping with $\beta(\gamma)=\gamma$,
whereas for rather
small ($+$) $D=0.6$, the data show a slight deviation from the asymptote
at least within the available $N=32$.
This deviation should be regarded as a transient behavior,
and for sufficiently large $N$, 
the multi-criticality 
$\dot{\nu}^{(\gamma)}=1$
(\ref{multi-critical_critical_exponent})
would be realized for
generic nonzero $D \ne 0$. 
As mentioned above, the $\dot{\nu}^{(\gamma)} =1$ result implies
$\phi =1/2$
(\ref{D_crossover_critical_exponent})
which differs $\phi=1$ (\ref{crossover_critical_exponent_D0})
for $D=0$
\cite{Nishiyama19}.
Therefore, 
to the extent of the undertaken parameter space,
only the multi-critical exponents, $\dot{\nu}^{(\gamma)}$ and $\phi$,
are affected by
the DM interaction, and the other features are retained unlike the one-dimensional magnet.

\section{\label{section3}
Summary and discussions}

The two-dimensional transverse-field $XY$ model (\ref{Hamiltonian_original})
with the DM interaction $D$
was investigated with the exact diagonalization method,
which enables us to treat
the complex-valued matrix elements due to $D$.
We implemented the screw-boundary condition
(\ref{Hamiltonian})
to the finite-size cluster with $N \le 32$ spins.
As a preliminary survey,
we analyzed the 
$H$-driven criticality 
at the Ising point, $\gamma=1$ and $D=0$.
With the fidelity susceptibility $\chi_H^{(H)}$ (\ref{H-driven_fidelity_susceptibility}),
we estimated the critical point as 
$H_c= 3.065(20)$ [Eq. (\ref{critical_point})], which agrees with the preceding estimates, 
$H_c=3.0442(4)$
\cite{Albuquerque10}
and 
$3.05(1)$
\cite{Henkel84},
confirming
that the order-disorder phase transition belongs to
the classical three-dimensional-Ising universality class, $x_{3DI}=1.1739$ (\ref{3DI_scaling_dimension}).
We then turn to the analysis of the $\gamma$-driven phase transition.
In order to cope with the level crossings \cite{Mukherjee11} at $\gamma=0$,
we evaluated the $\beta$ function, $\beta(\gamma)$ (\ref{beta_function}), and from its slope
$1/\nu^{(\gamma)}$
beside $\gamma \approx 0$, we obtained an estimate $\nu^{(\gamma)}=1/2$ (\ref{gamma-driven_critical_exponent})
for generic values of $D$ and $H$.
According to the real-space-renormalization-group analysis for $D =0$ and $H=0$ \cite{Usman16},
the estimates, $\nu^{(\gamma)}=1.14^{-1}$ and
$1.35^{-1}$, were obtained
from the peak position $\gamma_{max}$ and the peak height 
$\partial_\gamma C|_{\gamma=\gamma_{max}}$
of 
$\partial_\gamma C$
($C$: concurrence), respectively.
Likewise, the critical exponent $\nu^{(\gamma)}=0.4869$ \cite{Farajollahpour18} was obtained
from the energy gap for $D=0$ and $H=0$, and
it was claimed that
the exponent $\nu^{(\gamma)}(D)$ should be a monotonically decreasing function.
As mentioned above, our result indicates that the index $\nu^{(\gamma)}=1/2$ 
(\ref{gamma-driven_critical_exponent}) is robust
against $D$ and even $H$.
It was also determined how the critical exponent
$\nu^{(\gamma)}$ changes at the multi-critical point $H=H_c(0)$.
Our result indicates that the multi-criticality turns into
$\dot{\nu}^{(\gamma)}=1$
(\ref{multi-critical_critical_exponent})
for $D \ne 0$
eventually, and accordingly, the crossover critical exponent changes to
$\phi=1/2$ (\ref{D_crossover_critical_exponent}).
Therefore, the DM interaction alters the power-law singularity of the 
phase boundary,  
$H_c\sim |\gamma|^2$
(\ref{crossover_critical_exponent}), as shown in Fig. \ref{figure2} (c).

According to the spherical model analysis \cite{Wald15},
the phase boundary $H_c(\gamma)$ should exhibit a monotonic increase in large dimensions $d>2.065$,
whereas 
a reentrant behavior may occur in $d<2.065$.
It is thus expected that the multi-criticality is sensitive to the perturbations such as $D$, because
the concerned dimensionality $d=2$ locates around the marginal point $d=2.065$.
In fact,
our result indicates that the DM interaction alters the crossover exponent to $\phi=1/2$ (\ref{D_crossover_critical_exponent}).
It is tempting to consider the easy-plane SU$(N)$ magnet \cite{DEmidio16}
with the in-plane anisotropy
and the DM interaction
to see whether the phase boundary exhibits exotic features such as the reentrant behavior.
This problem is left for the future study.

\section*{Acknowledgment}

This work was supported by a Grant-in-Aid
for Scientific Research (C)
from Japan Society for the Promotion of Science
(Grant No.
20K03767).

\section*{References}


\end{document}